\documentclass[english,twocolumn,prl,superscriptaddress,longbibliography]{revtex4-1}
\usepackage[T1]{fontenc}
\usepackage[latin9]{inputenc}
\usepackage{color}
\usepackage{amsmath}
\usepackage{graphicx}

\usepackage[colorlinks=true,linkcolor=blue,urlcolor=blue,citecolor=blue,anchorcolor=blue]{hyperref}
\makeatletter
\usepackage{amsmath}
\usepackage{subfigure}
\usepackage{graphicx, mathtools}
\usepackage[normalem]{ulem}
\usepackage{soul}

\makeatother

\usepackage{babel}
\begin{document}
\newcommand{\be}{\begin{equation}}
\newcommand{\ee}{\end{equation}}
\newcommand{\bn}{\begin{eqnarray}}
\newcommand{\en}{\end{eqnarray}}
\newcommand{\ii}{\'{\i}}
\newcommand{\ca}{\c c\~a}
\newcommand{\uc}{\uppercase}
\newcommand{\tb}{\textbf}
\newcommand{\bw}{\begin{widetext}}
\newcommand{\ew}{\end{widetext}}

\title{ Quantum critical scaling of the conductivity tensor at the metal-insulator transition in Nb$_{1-x}$Ti$_{x}$N }

\author{D. Hazra}\email{iamdibyenduhazra@gmail.com}
\affiliation{Univ. Grenoble Alpes, CEA, INAC, PHELIQS, 38000 Grenoble, France}
\author{Prosenjit Haldar}\email{haldar@irsamc.ups-tlse.fr}
\affiliation{Laboratoire de Physique Th\'{e}orique, IRSAMC, Universit\'{e} de Toulouse, CNRS, UPS, France}
\affiliation{Centre for Condensed Matter Theory, Department of Physics, Indian Institute of Sciences, Bangalore 560012, India}
\author{M. S. Laad}\email{mslaad@imsc.res.in}
\affiliation{Institute of Mathematical Sciences, Taramani, Chennai 600113, India and\\
Homi Bhabha National Institute Training School Complex, Anushakti Nagar, Mumbai 400085, India}

\author{N. Tsavdaris}
 \affiliation{Univ. Grenoble Alpes, CNRS, Grenoble INP, SIMaP, 38000 Grenoble, France }
\author{A. Mukhtarova}
\affiliation{Univ. Grenoble Alpes, CEA, INAC, PHELIQS, 38000 Grenoble, France}
\author{M. Jacquemin}
 \affiliation{Univ. Grenoble Alpes, CNRS, Grenoble INP, SIMaP, 38000 Grenoble, France }
\author{ R. Albert}
\affiliation{Univ. Grenoble Alpes, CEA, INAC, PHELIQS, 38000 Grenoble, France}
\author{ F. Blanchet}
\affiliation{Univ. Grenoble Alpes, CEA, INAC, PHELIQS, 38000 Grenoble, France}
\author{ S. Jebari}
\affiliation{Univ. Grenoble Alpes, CEA, INAC, PHELIQS, 38000 Grenoble, France}
\author{ A. Grimm}
\affiliation{Univ. Grenoble Alpes, CEA, INAC, PHELIQS, 38000 Grenoble, France}

\author{E. Blanquet}
 \affiliation{Univ. Grenoble Alpes, CNRS, Grenoble INP, SIMaP, 38000 Grenoble, France }
\author{F. Mercier}
 \affiliation{Univ. Grenoble Alpes, CNRS, Grenoble INP, SIMaP, 38000 Grenoble, France }

\author{C. Chapelier}\email{claude.chapelier@cea.fr}
\affiliation{Univ. Grenoble Alpes, CEA, INAC, PHELIQS, 38000 Grenoble, France}
\author{M. Hofheinz}\email{max.hofheinz@usherbrooke.ca}\affiliation{Univ. Grenoble Alpes, CEA, INAC, PHELIQS, 38000 Grenoble, France}
\affiliation{Institut quantique and D\'{e}partement GEGI, Universit\'{e} de Sherbrooke, Sherbrooke, QC, Canada}
\author{Pratap Raychaudhuri}\email{praychaudhuri@tifr.res.in}\affiliation{DCMPMS, Tata Institute of Fundamental Research, Homi Bhabha Rd, Mumbai 400005, India}

\begin{abstract}
In contrast to the Landau paradigm, a metal-insulator transition (MIT), driven 
purely by competition between itinerance and localization and unaccompanied by any conventional (e.g, magnetic) order-disorder instabilities, admits no obvious 
local order parameter. Here, we present a detailed analysis of the quantum criticality in magneto-transport data on the alloy Nb$_{1-x}$Ti$_{x}$N across a Ti-doping-driven a MIT.  
We demonstrate, for the first time, clear and novel quantum criticality 
reflected in the {\it full} 
conductivity tensor across the MIT.  Wide ranging, comprehensive accord with 
recent theoretical predictions strongly suggests that these unanticipated findings are representative of a {\it continuous} MIT of the band-splitting type, 
rather than a conventional Anderson disorder or a ''pure'' correlation-driven 
first-order Mott type.    
\end{abstract}


\maketitle


Quantum phase transitions (QPT) between different phases continue to underpin novel developments in quantum matter~\cite{sachdev}.  Among QPTs, a metal-insulator transition (MIT), driven either dominantly by electron correlations~\cite{imada}, disorder~\cite{anderson,tvrRMP}, or both~\cite{kravchenko1996electric,sarachik}, is distinguished by lack of a Landauesque, local order parameter.  Such novel QPTs solely involve competition between kinetic energy-induced delocalization and correlation- and/or disorder-induced localization of carriers.  Interestingly~\cite{Kotliar_strange_metal}, recent work shows that such QPTs may underlie ''strange'' metallicity involving partial Mott localization of carriers.
Thus, investigation of ``Mott quantum criticality'' is a timely issue of great relevance to emergence of unusual electronic behavior in quantum matter. 

   In contrast to pure correlation-driven (first-order) cases, 
(strong or weak) disorder-driven MITs show genuine quantum criticality~\cite{gangoffour,sudip1997}.
   Notwithstanding non-perturbative interplay of itinerance and localization, unveiling quantum critical dynamics is facilitated by diverging spatio-temporal dynamical fluctuations as $T\rightarrow 0$ near the quantum critical point (QCP), permitting use of scaling relations to characterize finite $T, \omega$ (here, $\omega$ is the excitation energy) responses. Near a QPT, both, the spatial correlation length, $\xi_{x}$ and correlation time, $\tau$, diverge like $\xi_{x}\propto |x-x_{c}|^{-\nu}$ and $\tau\propto \xi_{x}^{z}\propto |x-x_{c}|^{-z\nu}$, with $x$ a tuning parameter, $x_{c}$ its critical value at the QCP, $\nu$ the correlation length exponent and $z$ the dynamical 
critical exponent.  A new thermal timescale, $t_{th}\simeq \hbar/k_{B}T$, viewable as the system size in the temporal direction, appears at finite $T$.  Thus, 
finite-$T$ data can be finite-size-scaled in terms of the ratio $t_{th}/\tau$ 
to track the growth of critical fluctuations as a system enters the quantum 
critical fan above a QCP at finite $T$.  Specifically, singular parts of 
physical response functions in the quantum critical region must be universal 
functions of $\tau/t_{th} \propto T/|x-x_{c}|^{z\nu}$.

At a continuous (here, achieved by Ti-substitution in NbN) MIT, the control parameter is the deviation of the doping from 
its critical value at the MIT, $x_{c}$, and physical responses must scale as $T/|x-x_{c}|^{z\nu}$ in the quantum critical regime of the MIT.  Moreover, one also expects 
deeper manifestations of quantum criticality: $(i)$ in a wide $x$-regime around $x_{c}$, $\tilde{\rho}_{xx}(T,\delta x)=1/\tilde{\rho}_{xx}(T,-\delta x)$, with  $\tilde{\rho}_{xx}=\rho_{xx}(T)/\rho_{c}(T)$ the scaled dc longitudinal resistivity, $\rho_{c}$ is the resistivity at $x=x_c$ and $\delta x=(x-x_{c})/x_{c}$
the distance from the QCP. This property, first seen in pioneering studies for the 2D electron gas in MOSFETS~\cite{scalingMosfet}, implies that the M- and I-phases are mapped into each other by a simple reflection, exposing a novel {\it duality} between them.  The upshot thereof is that the $\beta$- (or Gell-Mann Low function, $\beta=\frac{d\ln(\tilde{\rho}^{-1}_{xx})}{d\ln(L)}$,  $L$ is the length scale) has precisely the same form on both sides of the QCP~\cite{sudip1997},  $(ii)$ quantum critical scaling, {\it i.e}, 
the $\tilde{\rho}_{xx}(T)$ curves for both, insulator (I) and metal (M) phases {\it separately} collapse onto two universal curves when plotted versus a ``scaling 
variable'', $(T/T_{0})^{1/z\nu}$.  Here, $T_{0}(\delta x)= c|\delta x|^{z\nu}\propto \xi_{x}^{-z}$ (with c is a constant) vanishes at the QCP, reflecting a divergent localization length.  
 Thus, we expect that the $dc$ resistivity exhibits a scaling law: $\tilde{\rho}_{xx}(T, \delta x)=F_{\pm}(T/|\delta x|^{z\nu})$, with $F_{\pm}$ being a scaling functions in the M ($+$) and I $(-)$ phases, and $F_{\pm}$ constrained by the reflection symmetry.
 The exponents $z$ and $\nu$, extracted from such analysis, help identify whether the MIT belongs to the Mott, Anderson, or of an other, unconventional type.
 \begin{figure}
\includegraphics[width=1.0\linewidth]{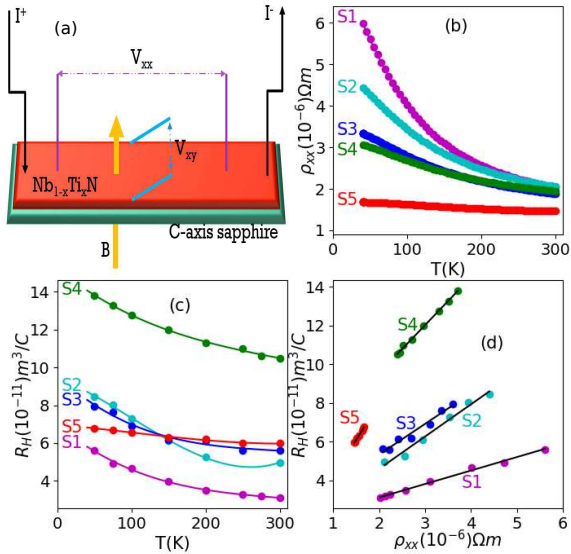}
\caption{(a) {\bf Schematic measurement geometry of Nb$_{1-x}$Ti$_{x}$N:} $\rho_{xx}$ and $\rho_{xy}$ are determined by measuring the longitudinal ($V_{xx}$) and Hall ($V_{xy}$) voltages, respectively, at a fixed d.c. bias current $I$. The magnetic field $B$ is applied perpendicular to the sample. (b) The temperature variation of $\rho_{xx}$ of all five samples at $B=0$.  $\rho_{xx}(T)$ decreases for all five samples with increasing $T$, affirming the ``Mooij correlation'' in the metallic phase. 
(c) Variation of $R_H$ as a function of temperature T. The solid lines are cubic polynomial fits. (d) Variation of $R_H$ as a function of $\rho_{xx}$. The solid lines are straight line fits.}
\label{fig:fig2}
\end{figure} 

{\bf Experiments and results --}  We now present our results for 
Nb$_{1-x}$Ti$_{x}$N, relegating the details in the {\bf SI1}. Five samples of Nb$_{1-x}$Ti$_x$N films of thickness $\sim$ 10 nm with different $x$ were grown on c-axis sapphire by high temperature chemical vapour deposition. All the films grow epitaxially as was evident from high resolution transmission electron microscopy and x-ray diffraction studies~\cite{tsavdaris2017hazra,hazra2018superconducting}. The relative composition of Ti and Nb were controlled by the gas flow rate. 
Fig.~\ref{fig:fig2}(a) shows a schematic of the set-up used to measure the longitudinal ($V_{xx}$) and Hall ($V_{xy}$) voltage, respectively, at a fixed d.c. bias current $I$ in presence of a magnetic field $B$, applied perpendicular to the sample. The measurements were carried out in a commercial Physical Property Measurement System (Quantum Design).  In Fig.~\ref{fig:fig2}(b), we show that $\rho_{xx}(T)$ at zero magnetic field decreases with T ($d\rho_{xx}/dT <0$), clearly revealing the ``Mooij correlation'' well into the metallic phase up to room temperature. $\rho_{xy}$ is determined by sweeping the magnetic field $B$ from 0 to 8 Tesla. At a fixed $T$, $\rho_{xy}$ varies linearly with $B$, giving the Hall coefficient, $R_{H}=\sigma_{xy}/B\sigma_{xx}^{2}$. In Fig.~\ref{fig:fig2}(c), we show that $R_{H}(T)$ decreases with $T$ in a way similar to $\rho_{xx}(T)$, as also found in an earlier work on NbN~\cite{madhavi}.  Indeed, as shown in Fig.~\ref{fig:fig2}(d),  $R_{H}(T)=C+C'(x)\rho_{xx}(T)$, and the ratio $r=\frac{\Delta R_{H}/R_{H}}{\Delta\rho_{xx}/\rho_{xx}}$ lies between $0.7$ and $0.9$. The disorder level in our samples is characterized by the Ioffe-Regel parameter $k_{F} \ell$ which is determined at 50 K in the standard way~\cite{madhavi}. 

\begin{figure}
\includegraphics[width=1.0\linewidth]{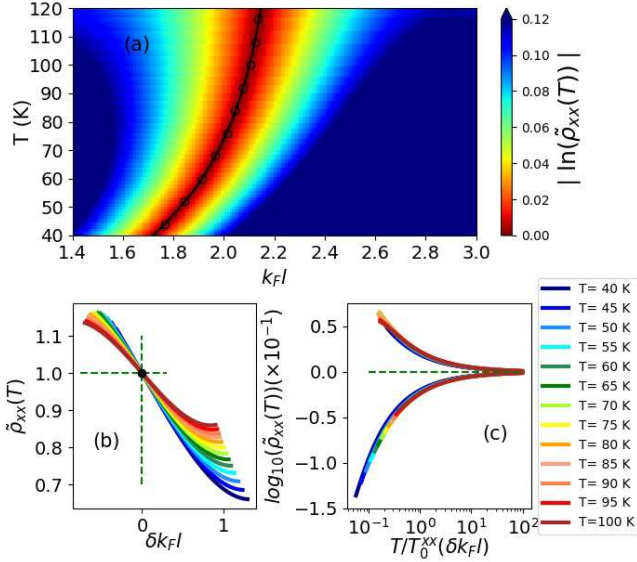}
\caption{Color plot of the T-dependent scaled dc resistivity as a function of $k_{F}l$ (upper panel (a)). The "fan-like" form expected for quantum criticality is clearly seen. Black empty circles represent the "quantum Widom line", defined as the crossover scale that describes the increasingly rapid crossover between a metal and an insulator as $T\rightarrow 0$. In Nb$_{1-x}$Ti$_{x}$N, the metallic side of the QPT is pre-empted by a low-{\it T} transition to superconductivity~\citep{hazra2018superconducting}. In the lower panels, ``Duality'' between the metal and insulator revealed in the dc longitudinal resisitivity: (a) Crossing point in the scaled resistivity $\tilde{\rho}_{xx}(T)=\rho_{xx}(T)/\rho_{c}(T)$ showing a clear MI ``transition'' at $\delta k_{F}l=k_{F}l-(k_{F}l)_{c}=0$, (b) $\tilde{\rho}_{xx}(T)$ vs $T/T^{xx}_0=T/c\mid\delta k_{F}l\mid^{(z\nu)_{xx}}$ exhibits clear quantum critical scaling, and mirror-symmetry, with $(z\nu)_{xx}\approx 1.51$.  } 
\label{fig:fig3}
\end{figure}

{\bf Quantum critical scaling --}  In order to test for signatures of quantum criticality, we perform the following analysis.  First,  we obtain $\rho_{xx}(T)$ on all points of $T$ vs $k_F l$ plane by polynomial fitting of order 3 with $T$ in the range $40-120~$K and $k_{F}l$ in the range $1.4-3.0$.  At a given temperature ($T$), we determine the metal-insulator critical value of $k_F l$, $(k_Fl)_c$, as the inflection point of $\rho_{xx}$ vs $k_F l$ curve. The line $\delta k_F l=k_F l-(k_F l)_c=0$ separates the metallic ($\delta k_F l>0$) and the insulating ($\delta k_F l<0$) phases at different temperatures.  As shown in Fig.~\ref{fig:fig3}(a), the scaled longitudinal resistivity, $\tilde{\rho}_{xx}(T,\delta k_F l)=\rho_{xx}(T,\delta k_F l)/\rho_{c}(T)$), (here $\rho_{c}(T)$ is the resistivity at $(k_F l)_c$), changes continuously from the metallic ($\delta k_F l>0$) to the insulating ($\delta k_F l<0$) phase. Following earlier procedure~\cite{scalingMosfet}, the horizontal axis is scaled to $\mid \delta k_{F}l\mid/t_{xx}(T)$, where $t_{xx}(T) \sim T^{1/(z\nu)_{xx}}$, such that all the metallic (M) and insulating (I) curves corresponding to different $T$ fall on one metallic and one insulating master curve, respectively.  This procedure allows extraction of the product $(z\nu)_{xx}$ via a fit to the functional form $t_{xx}(T)$ versus $T$. We use a similar protocol to unearth scaling of $\sigma_{xy}(T)$.

  Our results expose {\it all} the characteristic signatures of quantum-critical behavior expected at a continuous MIT.  In particular, in
Fig.~\ref{fig:fig3}(b), we show that $\rho_{xx}(T,\delta k_F l)$ exhibits a clean crossing point at $\delta k_{F}l=0$ when plotted as a function of $\delta k_{F}l$. Moreover, as shown in Fig.~\ref{fig:fig3}(c), clear quantum-critical scaling is obtained upon plotting log$(\rho_{xx}(T,\delta k_F l)/\rho_{c})$ versus $T/T^{xx}_{0}(\delta k_{F}l)$, with $T^{xx}_{0}=c\mid\delta k_{F}l\mid^{(z\nu)_{xx}}$ affirming the quantum critical character of the 
MIT.  We extract $(z\nu)_{xx}=1.5$, a value that substantially differs from $z\nu=0.67$~\cite{terletska} for a purely correlation driven Mott MIT, but in the range observed earlier for MOSFETs~\cite{scalingMosfet} ($z\nu=1.6$). 
More remarkably, we unearth 
clear ``mirror symmetry'' between metallic and insulating branch in Fig.~\ref{fig:fig3}(c). 

\begin{figure}
\includegraphics[width=1.0\linewidth]{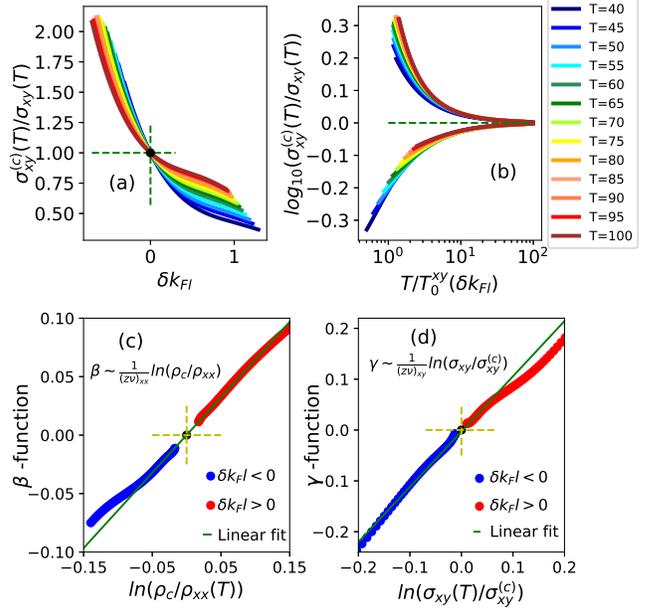}
\caption{(a) Crossing point in (scaled) Hall conductivity $\sigma^{(c)}_{xy}(T)/\sigma_{xy}(T)$ is further proof of the MIT at $\delta k_{F}l=k_{F}l-k_{F}l_{c}=0$, (b) Remarkably, $\sigma^{(c)}_{xy}(T)/\sigma_{xy}(T)$ vs $T/T^{xy}_0=T/c\mid\delta k_{F}l\mid^{(z\nu)_{xy}}$ shows 
quantum critical scaling and mirror-like symmetry with $(z\nu)_{xy}=0.97$, (c) $\beta$- (or Gell-Mann-Low) function for the full dc conductivity tensor: In the left panel, $\beta$-function vs $ln(g_{xx})$ and (d) in the right panel, $\gamma$-function vs $ln(g_{xy})$. Both $\beta$ and $\gamma$-functions show clear $ln(g)$ dependence, even quite well into the metallic phase, testifying 
to a very unusual manifestation of Mott-like quantum criticality.}
\label{fig:fig4}
\end{figure}

Remarkably, the off-diagonal conductivity $\sigma_{xy}(T, \delta k_Fl)$,  also exhibits  a similar and anomalous quantum critical scaling (see Fig.~\ref{fig:fig4}(a)). While well known for the longitudinal resistivity~\cite{sarachik}, such a clear signature of a MIT in the off-diagonal has never been seen to our best knowledge, though recent work on disordered TaN films~\cite{kapitulnik} hints such a possibility (but there as a function of $B$). Specifically, the scaled Hall conductivity, log$(\sigma_{xy}^{(c)}/\sigma_{xy}(T,\delta k_F l))$, for the M- and I-phases, separately coalesce onto two universal functions of $T/T^{xy}_{0}(\delta k_{F}l)$, precisely as for the dc resistivity. Moreover, in Fig.~\ref{fig:fig4}(b), we show that mirror symmetry is obeyed in this case as well.  Thus, the 
duality between the M- and I-phases also manifests in Hall conductivity. 

   Even more striking manifestations of the unusual quantum criticality are visible upon
extracting the $\beta$-function, defined as $\beta=\frac{d\ln(\rho_c/\rho_{xx}(T))}{d\ln(T)}$ with $L=T^{-1/z}$, with $z$ the dynamical exponent. We calculate $\beta$-function numerically from the Fig.~\ref{fig:fig3}(c). In Fig.~\ref{fig:fig4}(c), we show 
that $\beta(\rho_c/\rho_{xx}(T)) \simeq \log(\rho_c/\rho_{xx}(T))$ and is continuous across the MIT: 
while one expects $\rho_{xx}(T) \simeq$ exp$[(T/T_{0})^{-1/(z\nu)_{xx}}]$ in an insulating phase, it is remarkable that this behavior continues to hold quite deep in the metallic
 phase as well.  Separately, reflection symmetry is also obeyed very well, as we show in Fig.~\ref{fig:fig3}, confirming that the M- and I-phases are dual to each other. 
 A truly surprising finding of ours is that the $\gamma$-function (the Gell-Mann Low function for $\sigma_{xy}$, $\gamma=\frac{d\ln(\sigma_{xy}(T)/\sigma^{(c)}_{xy})}{d\ln(T)}$) also shows a completely unanticipated
 Mott-like scaling:
in Fig.~\ref{fig:fig4}(d), we show that $\gamma(\sigma_{xy}/\sigma^c_{xy})\simeq$ log$(\sigma_{xy}/\sigma^c_{xy})$ as well, even deep in the M-phase.  We can directly extract the exponent 
$z\nu$ from these, and find that $(z\nu)_{xx}=1.51$ and thus, $T_{0}^{(xx)}\simeq \mid\delta k_{F}l\mid^{1.51}$, while $(z\nu)_{xy}=0.97 \simeq 1.0$, implying that
$T_{0}^{(xy)}\simeq \mid\delta k_{F}l\mid^{1.0}$. These 
 distinct $(z\nu)$ values suggest that the decay of longitudinal and Hall currents are controlled by distinct relaxation rates.   
Indeed, this holds near the MIT : while $\rho_{xx}(T)=[\sigma_{0}+A T^{0.7}]^{-1}$ between $50$~K and $300$~K (shown in Fig. SI3 in the {\bf SI1}), it is obvious that the Hall angle ($\theta_{H}$) defined by {\it B}cot$\theta_{H}(T)=\rho_{xx}(T)/R_{H}(T)$, exhibits a different $T$-dependence (shown in Fig. SI5).  It immediately follows that the transverse relaxation rate, 
$\tau_{H}^{-1}\simeq$ cot$\theta_{H}$, is distinct from the longitudinal relaxation rate, $\tau^{-1}(T)\simeq \rho_{xx}(T)$, manifesting the two-relaxation rates scenario.  Such novel features  are signatures of a ``strange'' metal~\cite{cuprates}.  But this arises whenever $R_{H}(T)$ is sizably $T$-dependent, as in our case.  This manifests the non-perturbative breakdown of Landau fermion-like quasiparticles in the quantum critical region 
associated with a continuous MIT.  It is clearly {\it not} related to proximity to a
$T=0$ melting of any quasiclassical order, nor to any vagaries of a Fermi surface reconstruction, since no Fermi surface can possibly exist in the very bad 
metallic state close to the MIT (see the resistivity data in Fig.~\ref{fig:fig3}). 
It is truly remarkable that the full conductivity tensor nevertheless 
exhibits a novel,
 Mott-like quantum critical scaling at the MIT in Nb$_{1-x}$Ti$_{x}$N: this has 
 a range of deeper implications, detailed below.

{\bf Discussions and Conclusions --}   
To appreciate the novelty of our findings, we emphasize that our results 
contradict expectations from both, the weak localization (WL) view of an Anderson MIT as well as the correlation-driven Mott MIT.  In the first scenario, while scaling of $\sigma_{xy}$ is long known, semiclassical arguments in that case dictate that both, $\beta(\rho_c/\rho_{xx}(T))$ and $\gamma(\sigma_{xy}/\sigma^c_{xy})$ scale like $\beta(\rho_c/\rho_{xx}(T))\sim (D-2)-(\rho_{xx}/\rho_c)$, resulting in the 
quantum correction for $\sigma_{xy}$ being twice that for $1/\rho_{xx}$.  It 
turns out that this holds only as long as the inverse Hall constant, related to $h(L)\simeq L^{D-2}/R_{H}B$~\cite{shapiro}, scales {\it classically} as $L^{D-2}$ for small $B$.  Given our finding of a sizably $T$-dependent $R_{H}$, 
especially near the MIT, 
this assumption obviously breaks down in our case.  Additionally, we find $0.7 < r =\frac{\Delta R_{H}/R_{H}}{\Delta\rho_{xx}/\rho_{xx}}< 0.9$, in stark contrast to the prediction of a universal value $r=2.0$ in 
WL theory.  On the other hand, our 
results are also irreconcilable in a pure correlation driven Mott scenario: apart from
the fact that the MIT would have to be first order at low $T$ with a bad-metallic, linear-in-$T$ resistivity at the finite-$T$ critical point~\cite{terletska,kanoda}, the 
low-$T$ correlated 
metallic phase away from the critical point would be a heavy Landau Fermi 
liquid giving, for example, $\rho_{xx}(T)\simeq \rho_{0}+aT^{2}$.  Both are clearly in conflict with our finding of 
$(d\rho_{xx}/dT)<0$ (Mooij correlation) over a wide $T$-scale, $T_{c}<T<300$~K,
 well into the metallic side 
of the MIT. Moreover, in our finding the value $(z\nu)_{xx}=1.5$ substantially differs from $(z\nu)_{xx}=0.67$~\cite{terletska} for a purely correlation driven Mott MIT.  


Our findings raise the following fundamental issues: $(i)$ what is the nature of this novel QCP? and $(ii)$ what are the nature of the M- and I-phases?  Since the QCP we find is closer in nature to that seen in MOSFETs~\cite{scalingMosfet}, where $z\nu\simeq 1.6$, strong disorder (induced by Ti-doping) is dominant but interactions will also be important, especially 
near the MIT.  At a minimalist level, we propose an effective, random binary-alloy model as a simplest model of 
the real (random) alloy, wherein conduction $c$-fermions scatter off a random binary (since $Un_{id}=0,U$) disorder potential created by the localized $d$-fermions.  First-principles density-functional calculations show that the $d-p$ hybridization between the $d$-states (from $Nb, Ti$) and $p$-states (from $N$) is very weak in Nb$_{1-x}$Ti$_{x}$N~\cite{mathieu}.  We ignore it, especially since the strong, random $Un_{id}$ will generally quench this weak $d-p$ mixing.  This model shows a continuous MIT of the band-splitting type~\cite{freericks} even in DMFT as $U$ crosses a critical value $U_{c}$.  The lack of coherent hybridization rigorously precludes local Kondo screening, invalidating local Fermi liquidity
from the outset.  In the regime $k_{F}l\simeq O(1)$ of relevance here, DMFT and cluster-DMFT approaches yield a metallic state composed of a superposition of 
lower- and upper Hubbard bands, with progressive deepening of the charge pseudogap near the MIT~\cite{freericks,ourfirstpaper}.  In this situation,
transport is incoherent (since it solely involves transitions between the Hubbard bands), and the high-$T$ bad-metal behavior now persists down 
to $T=0$ (in fact, the only scale, as in local quantum critical scenarios, is
the temperature itself).  Our CDMFT studies for this model~\cite{qcmott,qchall} show clean quantum critical scaling of magneto-transport, with $(z\nu)_{xx}=1.31\simeq 4/3$ and $(z\nu)_{xy}=3/4$, comparing quite favorably with $(z\nu)_{xx}=1.51$ and $(z\nu)_{xy}=0.97$ found here. Furthermore, both $\rho_{xx}(T)$ and $R_{H}(T)$ increase with decreasing $T$ and roughly follow each other, precisely as predicted.  Also, our finding of $0.7<r<0.9$ is in close accord with $0.6<r<0.8$ from theory.
  Most importantly, both $\beta(g_{xx})$ and $\gamma(g_{xy})$ vary like log$y$ ($y=g_{xx},g_{xy}$).  This comprehensive accord thus strongly supports a band-coalescing MIT of the ''simplified Hubbard'' or ``binary alloy'' model type in our system.   Moreover, this also accords with expectations from a percolation-driven, continuous MIT expected in the strong-disorder limit of a binary-alloy disorder problem~\cite{fehske}.  Further, nanoscale electronic phase separation (EPS) is rigorously expected~\cite{kennedy} on theoretical grounds in binary-alloy models.  Our findings suggest that carrier dynamics occurs along percolative paths in such a strongly inhomogeneous background set by disorder. 

In Nb$_{1-x}$Ti$_{x}$N, as in NbN, a transition to superconductivity (SC) at very low 
$T$ prevents the observation of the QPT as $T\rightarrow 0$.  Thus, it is not possible to monitor
$\sigma_{xx}(\delta k_{F}l,T\rightarrow 0)=Ce^{2}/\hbar\xi_{x}\simeq (\delta k_{F}l)^{(\nu)_{xx}}$, to extract $(\nu)_{xx}$ and $(z)_{xx}$ separately.  Studying the electric-field $(E)$-driven MIT~\cite{sondhi} should resolve this issue: this is because, at low $T$, $\rho_{xx}(E,\delta k_{F}l)$ would depend only on $\frac{\delta k_{F}l}{E^{1/(z+1)\nu}}$, since the electric field introduces a new length scale, $L_{E}\simeq E^{-1/(z+1)}$, and as long as this is smaller than $L_{T}\simeq T^{-1/z}$, $E$-field scaling will obtain even in presence of heating effects~\cite{sondhi}.  This is clearly a direction for future experiments.

   Our findings strongly link the superconductor-insulator transition (SIT) in Nb$_{1-x}$Ti$_{x}$N (maybe also in TaN~\cite{kapitulnik}) to an underlying QCP associated with a {\it fermionic} MIT.  This has deeper implications for the nature of the SC instability, implying the need to go beyond purely bosonic descriptions of the SIT by incorporating critical fermionic dynamics at a fermionic MIT seen here.  Since the very bad metal is associated with a finite residual entropy, at least in quasi-local approaches, it seems that SC emerges as the {\it only} coherence-restoring instability (since competing instabilities to Wigner crystal/charge-density-wave, etc. will be inhibited by strong disorder-induced nanoscale inhomogeneity) that can quench this entropy as 
$T\rightarrow 0$.  In the critical regime
($k_{F}l\simeq 1.0$), such a SC must have a short pair-coherence length, $\xi_{pair}\simeq l\simeq O(a)$, with $a$ the lattice spacing and,
in fact, percolative dynamics in a nanoscale EPS state will also imply a strong phase fluctuation dominated SIT~\cite{madhavi2}.  Our work suggests that the non-trivial interplay between such critical dynamical fluctuations associated with the fermionic MIT, and onset of two-particle pair coherence in the SC phase (on the metallic side) is 
a crucial controlling factor influencing the nature of the SIT itself in such systems, and mandates incorporating this link into extant theories of the SIT, at least for such systems.


%

\end{document}